\documentclass[twocolumn,showpacs,prb,floatfix]{revtex4}
\usepackage{epsfig}
\usepackage{tabularx}

\newcommand{\be}{\begin{equation}}
\newcommand{\ee}{\end{equation}}
\newcommand{\beqn}{\begin{eqnarray}}
\newcommand{\eeqn}{\end{eqnarray}}

\begin{document}

\title{Reversal-field memory in magnetic hysteresis}
\author{H.~G.~Katzgraber$^1$, F.~P\'azm\'andi$^1$, C.~R.~Pike$^2$, 
Kai Liu$^1$, R.~T.~Scalettar$^1$, K.~L.~Verosub$^2$, G.~T.~Zim\'anyi$^1$}
\affiliation{$^1$Department of Physics, University of California, Davis,
California 95616\\ $^2$Department of Geology, University of California, Davis, 
California 95616}

\date{\today}

\begin{abstract}
We report results demonstrating a singularity in the 
hysteresis of magnetic materials, the reversal-field memory effect. 
This effect creates a nonanalyticity in the magnetization curves at a 
particular point related to the history of the sample. The microscopic 
origin of the  effect is associated with a local spin-reversal symmetry 
of the underlying Hamiltonian. We show that the presence or absence
of reversal-field memory distinguishes two widely studied models 
of spin glasses (random magnets).
\end{abstract}

\pacs{75.50.Lk, 75.40.Mg, 05.50.+q}
\maketitle

\section{Introduction}
\label{intro}

Hysteresis is a widely observed phenomenon.\cite{bertotti:98}
While many basic features are well 
understood\cite{sethna:93,lyuksyutov:99,zhu:90} a number of effects remain
unresolved, such as the noise spectrum,\cite{bertram:94}
frequency dependence,\cite{lederman:94} and exchange
bias\cite{meiklejohn:57,nogues:99} in the hysteresis of magnetic
recording media.

Here we report a novel memory effect in magnetic
systems that emerges when the magnetic field is first
decreased from its saturation value and then increased again 
from some reversal field $H_R$. 
We find that the system exhibits a singularity at the negative of the 
reversal field, $-H_R$, in the form of a sharp kink in the magnetization. 
Microscopically a finite number of ``symmetric clusters'' are required
for this effect to take place. In these clusters the central spins flip 
{\it after} all spins on the cluster boundary have flipped. In addition,
the central spins  experience zero effective local field so they are
symmetric with respect to the change of direction of the external field. 
We have observed this ``reversal-field memory'' effect numerically 
in spin glasses, as well as experimentally in magnetic 
thin films commonly used in recording media. 

To better illustrate the reversal-field memory effect we use the recently
introduced first order reversal curve (FORC) diagram 
approach\cite{pike:99} which captures the
distribution of characteristic properties of hysteretic systems in 
great detail. 
\vspace*{-0.4cm}
\section{Model and Algorithm}
\label{model}

We study the Edwards--Anderson (EA) Ising spin-glass 
Hamiltonian\cite{binder:86}
\begin{equation}
{\cal H}= \sum_{\langle i,j \rangle} J_{ij}S_iS_j -
                  H  \sum_i S_i \,. 
\label{eq:hamilton}
\end{equation}
Here $S_i = \pm 1$ are Ising spins on a square lattice of size 
$N = L \times L$ in two dimensions with periodic boundary conditions.
The exchange couplings $J_{ij}$ are random nearest-neighbor interactions 
chosen according to a Gaussian distribution with zero mean 
and standard deviation unity, and $H$ is the external magnetic field.
The complex energy landscape of the EA spin glass will give rise to
hysteretic behavior.

We simulate the zero temperature dynamics of the EA model
by changing the external field $H$ in small steps, first downward 
from positive saturation, and then upward from a reversal field $H_R$.
After each field step, the effective local field $h_i$ of each
spin $S_i$ is calculated:
\begin{equation}
h_i=\sum_{j} J_{ij}S_j - H \; .
\label{eq:local_field}
\end{equation}
A spin is unstable if $h_i S_i < 0$. We then flip a randomly chosen 
unstable spin and update the local fields at neighboring sites 
and repeat this procedure until all spins are stable.
\vspace*{-0.4cm}
\section{Results}
\label{results}

\begin{figure}
\epsfxsize=7.0cm
\begin{center}
\epsfbox{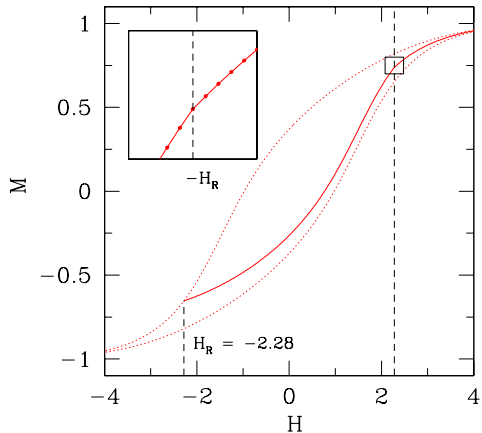}
\end{center}
\vspace*{-1.3cm}
\caption{\label{fig:kink}
Reversal-field memory effect:
starting from positive saturation, the field is decreased to a
negative reversal field $H_R$, then increased again. The inset
shows the kink at $-H_R$. Data are from a two-dimensional
EA Ising spin glass with $100^2$ spins.
The dotted line represents the major hysteresis loop, the dashed lines mark
$\pm H_R$, respectively.
In all figures the error bars are smaller than the symbols
and the data are averaged over 1000 disorder realizations.
In addition, $H$ and $M$ are in units of the interaction bond standard
deviation, 
which in this work is set to unity.
}
\end{figure}

Figure \ref{fig:kink} shows a typical reversal curve. The area 
around $-H_R$ is enlarged in the inset 
and shows a sharp ``kink''. The observation of any such sharp
feature in a disordered system, especially of finite size and after 
disorder averaging, is quite remarkable.

\begin{figure}
\epsfxsize=7.0cm
\begin{center}
\epsfbox{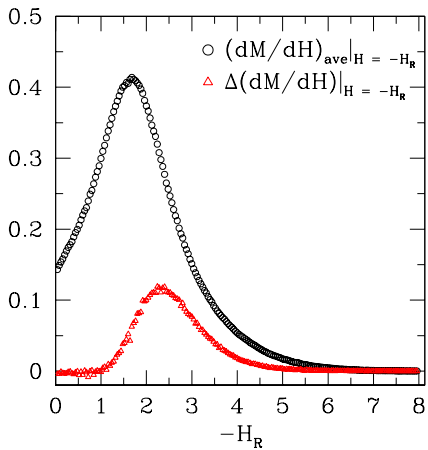}
\end{center}
\vspace*{-1.3cm}
\caption{\label{fig:delta}
The difference (triangles) and average (circles) of the left and right
derivatives on either side of $-H_R$ for the EA Ising spin glass.}
\end{figure}

The size of the change in slope can be characterized by
measuring the slope of the magnetization curves to the left and 
right of $-H_R$, and comparing the difference with the average 
(see Fig.~\ref{fig:delta}). The 
slope changes abruptly by as much as 30 \% as we pass through $H=-H_R$.
For this model, the range of reversal-field values for which the kink 
is present is roughly $-4.5 < H_{R} < -1.5$.

To develop an initial understanding of the microscopic origin of
reversal-field memory, we first describe the conditions under
which the effect occurs within the Preisach model.\cite{preisach:35}
In that approach, a magnetic system is described as
a collection of independent two-state ($\pm 1$)
switching units (``hysterons'').
Unlike Ising spins, which always align with their local field, 
the hysteron's state changes from $-1$ to $+1$ at a field
$H_b+H_c$ different from the field
$H_b-H_c$ required to switch the hysteron from $+1$ to $-1$.
Here $H_b$ is the bias field and $H_c$ is the coercivity of the hysteron.
Different systems are then distinguished by their distribution 
$\rho(H_b, H_c)$ of hysterons of a 
given bias and coercivity, the so-called ``Preisach function.''  

For ``symmetric hysterons,'' i.e., $H_b = 0$,
starting from a fully `up' polarized state and decreasing the field 
to a negative $H_R$ switches down all hysterons with $H_c<|H_R|$. 
Reversing the direction of the sweep and increasing
the field from $H_R$ to $-H_R$ switches back every switched hysteron. 
Thus at $H=-H_R$ saturation is reached, 
creating a kink in the magnetization.  
Symmetric hysterons therefore give rise to reversal-field 
memory, but only if their number is macroscopic. In this case
$\rho(H_b, H_c)$ must have a Dirac delta singularity at 
$H_b=0$ and $H_c=|H_R|$. If the kink is observable in a range of $H_R$ values, 
as in Fig.~\ref{fig:delta},
then the singularities of the Preisach function form 
a ridge along the $H_b=0$ axis for that range of $H_c$ values.

The spins in the EA model are not independent. For them to
behave as symmetric hysterons, they must possess a 
local spin-reversal symmetry. 
By local spin-reversal symmetry we mean that
the local field $h_i$, felt by $S_i$ [Eq.~(\ref{eq:local_field})], is
perfectly
reversed if the external field $H$ is reversed and all spins coupled to $S_i$
are reversed as well. In the EA Hamiltonian a cluster has this symmetry when
all neighbors of $S_i$ align with the local field before $S_i$ does,
both
for decreasing and increasing fields, i.e., $S_i$ flips down only 
after all of its neighbors are already negative,
and during the reverse sweep $S_i$ switches up only after all the neighbors.

The importance of the symmetric clusters can be appreciated by noticing that
every spin of the EA model seems to have local spin-reversal symmetry.
However, note that the spin configurations in general depend on the history of
the sample in the glassy phase.  Therefore, at $-H_R$ the neighbors of 
most spins do not necessarily point in a direction opposite of 
their direction at $H_R$,
thus most EA spins do not belong to symmetric clusters. 
Hence the model Hamiltonian possessing a local spin-reversal
symmetry is a necessary but not sufficient condition of having
symmetric clusters. In order to test this conclusion we studied
the random field Ising model (RFIM).\cite{binder:86,sethna:93} In this model 
the couplings $J_{ij}$ are chosen to be uniform, and the
disorder is introduced through random local fields $h_i$. Direct inspection
reveals that the RFIM does not possess a local spin-reversal symmetry,
nor did we find a magnetization kink in our simulations.

\begin{figure}
\epsfxsize=7.0cm
\begin{center}
\epsfbox{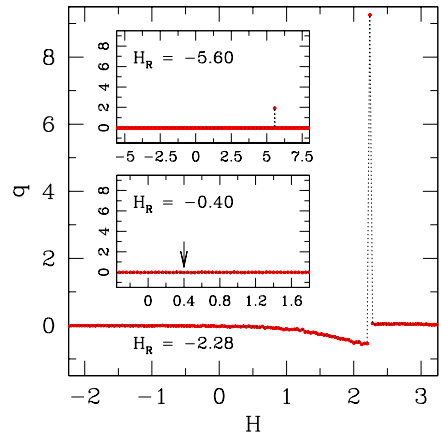}
\end{center}
\vspace*{-1.3cm}
\caption{\label{fig:overlap}
Overlap function $q$ of the spins flipping at $H_R$ and at $H > H_R$,
for $H_R = -2.28$. The insets show data for $H_R = -0.40$ and
$H_R = -5.60$, respectively. The arrow in the inset marks $H = 0.40$.
}
\end{figure}

To further illustrate the reversal-field memory effect,
we define an overlap function $q$ between the spins which flip at $H_R$
and the spins which flip at $H > H_R$:
\begin{eqnarray}
q(H) &=& \frac{1}{4}\sum_i[S_i(H_R + \delta) - S_i(H_R)] \times \nonumber \\
      &&\;\;\;\;\;\;\;\;\;\;\;\;\;\;\;\;\;[S_i(H + \delta) - S_i(H)] \; .
\label{overlap}
\end{eqnarray}
Here $\delta$ is the field step. In Fig.~\ref{fig:overlap} we show the overlap
versus $H > H_R$ for $H_R = -2.28$. Clearly
the spins which flip at $H_R$ are highly correlated with the
spins which flip at $-H_R$. 
This in turn shows that we have a macroscopic number of
symmetric clusters.
The insets show a much weaker effect at
$H_R = -0.40$ and $H_R = -5.60$, values outside the ridge of
Fig.~\ref{fig:delta}.

The reversal-field memory can be characterized in greater detail
by adapting a new tool developed for analyzing experimental
data for hysteretic systems.\cite{pike:99}
In this method, a family of FORCs 
is generated by varying $H_R$ in small steps. 
Let us denote by $M(H, H_R)$ the magnetization as the function of the applied
and reversal fields.  Computing the mixed second derivative with respect to
$H$ and the reversal field $H_R$ yields the ``FORC distribution''
$\rho(H_c, H_b)$, where $H_c=(H-H_R)/2$ represents the local coercivity and
$H_b=(H+H_R)/2$ the bias. FORC distributions are more general than Preisach
distributions since they are  
model independent and allow for asymmetries and negative regions. 

\begin{figure}
\epsfxsize=12cm
\begin{center}
\epsfbox{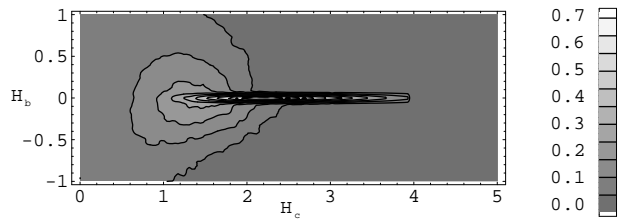}
\end{center}
\vspace*{-0.5cm}
\caption{\label{fig:forc-ea}
FORC diagram of the EA Ising spin glass. Note the ridge
along the $H_c$ axis.
}
\end{figure}

\begin{figure}
\epsfxsize=8cm
\begin{center}
\epsfbox{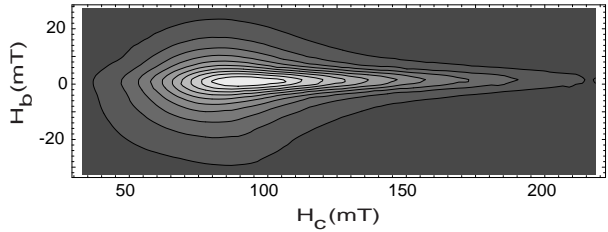}
\end{center}
\vspace*{-0.6cm}
\caption{\label{fig:kodak}
Experimental FORC diagram of a Kodak sample. Note the similarity
to the FORC diagram of the EA Ising spin glass shown in
Fig.~\ref{fig:forc-ea}.
}
\end{figure}

Figure \ref{fig:forc-ea} shows the FORC diagram of the EA model.
The ridge along the $H_c$ axis corresponds to the peak of 
Fig.~\ref{fig:delta} and is therefore a striking representation of the 
reversal-field memory effect. In contrast, the RFIM exhibits a vertical
structure, even though both models have quite similar major hysteresis loops.
This sensitivity of the FORC method makes it uniquely suited for 
analyzing details of hysteretic systems.

To demonstrate the existence of the reversal-field memory in magnetic media,
in Fig.~\ref{fig:kodak} we show an experimentally obtained FORC diagram
of well-dispersed 
noninteracting single-domain magnetic Co-$\gamma$-Fe$_2$O$_3$ particles.
Typically FORC diagrams of various systems exhibit extensive differences
(see for example Ref.~\onlinecite{pike:01a}).
There is a striking similarity between the experimentally and numerically
determined FORC diagrams: both exhibit a narrow ridge along the $H_c$ axis,
the ridge ``melting'' with decreasing $H_c$ and narrowing with 
increasing $H_c$. Simulations on more realistic systems with dipolar
interactions, such as the ones found in the experimental sample, also show
evidence of a ridge and therefore reversal-field memory.
\vspace*{-0.7cm}
\section{Conclusions}
\label{conclusions}

In conclusion, we report a reversal-field memory effect in disordered magnetic
systems that manifests itself as a sharp kink in first order
reversal curves,
and as a sharp ridge on the zero bias axis of FORC diagrams.
The effect has been observed numerically in the Edwards--Anderson Ising spin
glass and
experimentally in well-dispersed recording media samples.
The origin of this effect is
tied to a local spin-reversal symmetry, and has its microscopic origin in the
development of a macroscopic density of symmetric clusters.
Our studies also establish that the FORC method is a powerful diagnostic tool
for characterizing magnetic materials.
\vspace*{-0.4cm}
\section{Acknowledgments}
\label{acks}

This work was supported by NSF Grant Nos.~DMR-9985978, 99-09468, and
INT-9720440.

\bibliography{refs}

\end{document}